\documentstyle[prl,aps,twocolumn,psfig]{revtex}

\def\beq{\begin{equation}}
\def\eeq{\end{equation}}
\def\bea{\begin{eqnarray}}
\def\eea{\end{eqnarray}}
\draft

\begin{document}

\twocolumn[\hsize\textwidth\columnwidth\hsize\csname @twocolumnfalse\endcsname
\title{Nishimori point in the 2D $\pm J$ random-bond Ising model}

\author{A.\ Honecker$^{1}$, M.\ Picco$^{2}$ and  P.\ Pujol$^{3}$}
\address{$^{1}$Institut f\"ur Theoretische Physik, TU Braunschweig,
    Mendelssohnstr.\ 3, 38106 Braunschweig, Germany.}
\address{$^{2}$LPTHE\cite{CNRS1}, Universit\'e Pierre et Marie Curie, Paris VI,
        Universit\'e Denis Diderot, Paris VII\\
        Boite 126, Tour 16, 1$^{\it er}$ \'etage, 4 place Jussieu,
        75252 Paris C\'edex 05, France.}
\address{$^{3}$Laboratoire de Physique\cite{CNRS2},
    Groupe de Physique Th\'eorique, ENS Lyon,
    46 All\'ee d'Italie, 69364 Lyon C\'edex 07, France.}
\date{October 9, 2000; revised April 24, 2001}
\maketitle

\begin{abstract}
We study the universality class of the Nishimori point in
the 2D $\pm J$ random-bond Ising model by means of the numerical
transfer-matrix method. Using the domain-wall free-energy, we locate the
position of the fixed point along the Nishimori line at the critical
concentration value $p_c = 0.1094 \pm 0.0002$ and estimate $\nu = 1.33 \pm
0.03$. Then, we obtain the exponents for the moments of the spin-spin
correlation functions as well as the value for the central charge
$c = 0.464 \pm 0.004$.
The main qualitative result is the fact that percolation is now excluded as a
candidate for describing the universality class of this fixed point.
\end{abstract}

]

In the past years, the subject of disordered systems has known a huge
renewal of interest in the condensed matter and statistical mechanics
community. Among these disordered models, two-dimensional systems are of a
particular interest. Since the discovery of the unitary series of conformal
field theory (CFT) in 1984 \cite{BPZ}, exact values for the exponents of many
well known models of statistical mechanics have been given. However, an
equivalent classification for universality classes of such systems in the
presence of impurities is still missing.  A first big step towards a more
general classification has been done recently as a random matrices
classification \cite{Z}. The data for critical exponents in most of the
experimental relevant fixed points for impure system is however still not
available.

The Ising model on a square lattice is one of the most popular
two-dimensional systems. It is specified by the energy of a spin
configuration
\begin{equation}
E(\{S_i\}) = \sum_{\langle i,j \rangle} J_{i,j} \delta_{S_i,S_j}  \, ,
\label{RBI}
\end{equation}
where the sum is over all bonds and the coupling constants $J_{i,j}$
are bond dependent. We consider here the $J_{i,j}=\pm 1$ Random-Bond Ising
Model (RBIM) with the following probability distribution:
\begin{equation}
P(J_{i,j}) = p \delta(J_{i,j}-1) + (1-p) \delta(J_{i,j}+1) \, .
\label{pmJprob}
\end{equation}
Note that with these conventions, the pure model ($p=0$) is characterized
by $J_{i,j} = -1$ and thus has a ferromagnetic groundstate.

The RBIM is similar to other relevant disordered models such
as the Chalker-Coddington random network model which was proposed
originally in the context of the quantum Hall effect plateau
transition \cite{CF}. However, it is important to stress that
these systems have a different phase diagram and therefore
their fixed points have no reason to be in the same universality
class \cite{CRKHAL}.

The topology of the phase diagram of the RBIM depends crucially on the type
of disorder one considers. An instructive example is provided by a disorder
having only two possible values for the bonds with equal signs and
probabilities. It is by now well established \cite{JC} that the only non-trivial
fixed points are located at the extrema of the boundary of the
ferromagnetic phase, corresponding
to the pure Ising fixed point and a zero temperature fixed point which
turns out to be in the percolation universality class. It is interesting to
notice that percolation is also the universality class of the so-called
spin quantum Hall model \cite{GRLC}, another random network model.

\begin{figure}[ht]
\centerline{\psfig{file=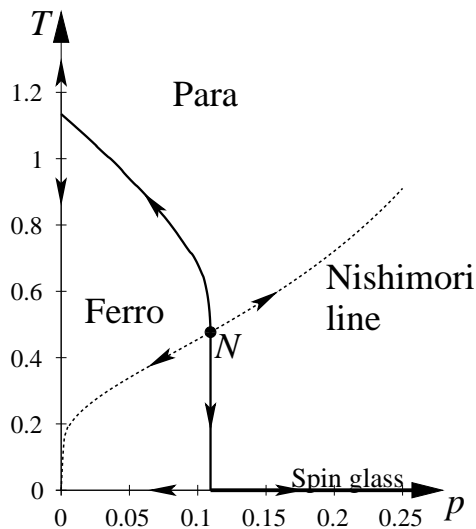,width=0.8\columnwidth}}
\caption{
Phase diagram of the two-dimensional $\pm J$ random-bond Ising model.
\label{phaseRBI}
}
\end{figure}

When the distribution contains also bonds with different signs (like in
(\ref{pmJprob})), the situation is more subtle. For a certain class of
probability distributions, Nishimori has shown that a so-called `Nishimori'
line exists where many properties can be calculated exactly \cite{N}.  For
the probability distribution (\ref{pmJprob}), this line is given by
\begin{equation}
{\rm e}^{\beta} =  {1-p \over p} \, ,
\label{NishimoriLine}
\end{equation}
with $\beta = 1/T$. On the Nishimori line, the internal energy can be
calculated exactly and an upper bound can be given for the specific
heat. Also of
interest is an equality of the moments of the spin correlation
functions (see below). Nishimori has further proven inequalities for the
correlation functions which yield important constraints on the topology of
the phase diagram which is shown in Fig.\ \ref{phaseRBI} for the
$\pm J$ RBIM \cite{footnote}. Since the Nishimori line is also invariant under
Renormalization Group (RG) transformations \cite{LG}, the intersection of
the Nishimori line and the Ferro-Para transition line must be a fixed
point. This so-called Nishimori point ($N$) corresponds to a new
universality class belonging precisely to the family of strong disorder
fixed points. The bold line in the phase diagram Fig.\ \ref{phaseRBI} is
the phase boundary between the
ferromagnetic and paramagnetic regions. At zero temperature the model has a
spin glass phase \cite{SG}. The three non-trivial fixed points along the
bold line are the pure Ising fixed point, the Nishimori point at the
crossing with the Nishimori line (dotted line) and the zero temperature
point, separating the ferro and the spin glass phases. The properties of
the latter point are still mostly unknown.  This point will be the subject of future
investigations \cite{prep}.

In the last years, many numerical and analytical efforts have been made in
order to identify the universality class of the Nishimori point. Very
recently, an analytic approach suggested that it is governed by an
$Osp(2n+1 \vert 2n) $ symmetry (and maybe $Osp(2n+2 \vert 2n)$ \cite{GRL},
but unfortunately, the classification of CFTs with such symmetries is still
missing. From a numerical point of view, there is by now a long list of
results \cite{MM,ON,SA,OI,AQdS}. An important observation is that all the
numerical results for the critical exponents tend to suggest that this
point is in the percolation universality class.  Because of the importance
of the statistical model on its own and its relevance for understanding
the plateau transition in the quantum Hall effect, it is crucial to
elucidate the similarity to percolation and the relation to the
super-symmetric CFT proposed in the literature.

In this letter we provide results of extensive numerical transfer-matrix
calculations of the Nishimori point with the binary distribution (\ref{pmJprob})
for bonds on the square lattice. We use the domain-wall free-energy to
accurately locate the critical concentration of disorder $p_c$ and to estimate
the exponent $\nu$. Then, we analyze the spin correlation functions and the
scaling of the free energy, giving accurate and novel results for the magnetic
exponent $\eta$ and central charge $c$. Apart from improving the identification
of this universality class (providing in particular values for the
central charge which have never been measured before), our main result is
that percolation is excluded as a possible candidate for describing this
fixed point.

First, we use the free energy of a domain wall \cite{MM} to locate the
critical point. For a strip of width $L$ the domain-wall free-energy $d_L$
is defined as \cite{footnDL}
\beq
d_L = L^2 \left(f_L^{(p)} - f_L^{(a)}\right) \, ,
\label{defDW}
\eeq
where $f_L^{(p)}$ is the free energy {\it per site} of a strip of width $L$
with {\it periodic} boundary conditions and $f_L^{(a)}$ the corresponding
one with {\it antiperiodic} boundary conditions.  $d_L$ is an observable
which can be used directly to study the RG flow under scale
transformations. In particular, it is constant at a fixed point.

We have computed $f_L^{(p)} = {\ln Z^{(p)} \over L N}$ and
$f_L^{(a)} = {\ln Z^{(a)} \over L N}$ employing a standard transfer
matrix technique with sparse matrix factorization (see, e.g., \cite{Night})
on strips of length $N = 10^6$. Since randomness is strong, care must be taken
to reduce fluctuations even if the free energies are
self-averaging. Therefore, we have fixed the concentration of bonds $p$
globally on a sample and computed $f_L^{(p)}$ and $f_L^{(a)}$ on the same
sample. Still, one needs around 1000 to 4000 samples of $L \times 10^6$
strips to obtain sufficiently small error bars for $L \le 12$. Even on
modern computers this needs an amount of CPU time which precludes the
analysis of wider strips. However, since we are looking for a fixed point,
no crossover effects are expected and it is legitimate to use small system
sizes.

\begin{figure}[ht]
\centerline{\psfig{file=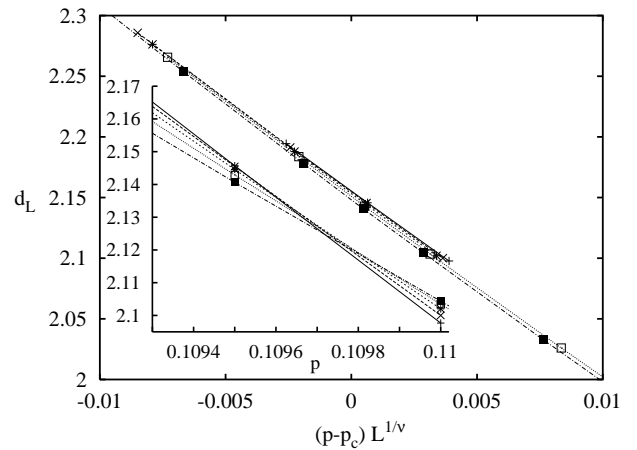,width=\columnwidth,angle=270}}
\smallskip
\caption{
Domain-wall free-energy. The inset shows the raw data and the
main panel the scaling collapse with $p_c = 0.1094$ and
$\nu = 1.33$. The symbols are for $L=8$ (filled boxes), $L=9$
(open boxes), $L=10$ ($*$), $L=11$ ($\times$) and $L=12$
($+$). Error bars are much smaller than the size of symbols.
\label{figScaleD}
}
\end{figure}

The inset of Fig.\ \ref{figScaleD} shows $d_L(p)$
along the Nishimori line (\ref{NishimoriLine}) in the vicinity
of the critical concentration $p_c$. A finite-size estimate for
$p_c$ is given by the crossing points $d_{L_1}(p_c) = d_{L_2}(p_c)$.
After extrapolation to an infinitely wide strip (details will
be given elsewhere \cite{prep}), one obtains
\beq
p_c = 0.1094 \pm 0.0002 \; .
\label{pcrit}
\eeq
This estimate improves upon the accuracy of earlier estimates
\cite{ON,SA,OI,AQdS}. It agrees perfectly with the transfer matrix
computations \cite{ON,AQdS} while we find a slightly smaller value
of $p_c$ than \cite{SA,OI}. We would like to mention that
(\ref{pcrit}) is confirmed by standard Monte Carlo simulations on
systems up to $32 \times 32$ sites \cite{prep} -- the present
estimate is just more accurate.

One can extract also the correlation length exponent $\nu$ from
$d_L$ if one assumes the scaling form
\beq
d_L(p-p_c) = d\left((p-p_c) L^{1/\nu} \right) \; .
\label{dScal}
\eeq
Again, we omit details \cite{prep} and quote just the final result
\beq
\nu = 1.33 \pm 0.03 \; .
\label{nuVal}
\eeq
The main panel of Fig.\ \ref{figScaleD} demonstrates that $d_L$ follows
indeed the scaling form (\ref{dScal}) with the parameters (\ref{pcrit})
and (\ref{nuVal}).

The result (\ref{nuVal}) is in complete agreement with $\nu = 1.32 \pm 0.08$
obtained by high-temperature series \cite{SA} as well as the value
$\nu = 4/3$ for percolation (see, e.g., \cite{StAh}).

Another important quantity is the magnetic exponent $\eta$. This exponent can
be measured, for example, by computing spin-spin correlation functions. As we
mentioned in the introduction, all along the Nishimori line the moments of
these correlation functions are equal two by two:
\beq
[\langle S(x_1,y_1) S(x_2,y_2)\rangle^{2k-1}]
= [\langle S(x_1,y_1) S(x_2, y_2)\rangle^{2k}]
\label{2by2}
\eeq
for any integer $k$. Here $[\cdots]$ stands for the average over the
disorder. Assume now that the correlation functions (\ref{2by2}) decay
algebraically on a plane and define by $x,y$ the coordinates on the infinite
cylinder of circumference $L$, with $x\in [1,L]$ and $y\in ]-\infty,+\infty[$.
Using a conformal mapping, one infers then the following behavior of the
correlation functions on the cylinder:
\beq
[\langle S(x_1,y) S(x_2,y)\rangle ^n] \propto
  \left(\sin\left({\pi (x_2-x_1)\over L}\right) L\right)^{- \eta_{n}} \, .
\label{fitsscf}
\eeq
For a pure system, one has $\eta_n = n \times \eta$.
On the other hand, in the case of percolation over Ising clusters, it is
easy to see that the moments of spin correlation functions are all
equal (and not only two by two). Then, if the Nishimori point is in the
percolation universality class, the exponents for the correlation
functions in (\ref{fitsscf}) should collapse to a unique value
$\eta_n = \eta$ at the critical point.

\begin{figure}[ht]
\centerline{\psfig{file=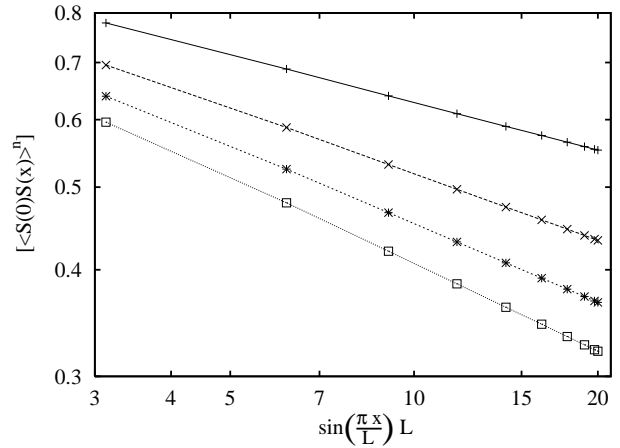,width=\columnwidth,angle=270}}
\smallskip
\caption{Moments of the spin-spin correlation function
for $p=0.1095$ and $L=20$. We only show the odd moments:
$n=1$ ($+$), $n=3$ ($\times$), $n=5$ ($*$) and $n=7$
(open boxes). Error bars are smaller than the size of the
symbols. The values of the exponents are given in (\ref{valExp}).
\label{ss}
}
\end{figure}

In order to verify this we have calculated the spin-spin correlation
functions on cylinders of width $L$ and length $400\times L$, ({\it i.e.}\
with the length $\gg L$) for $L$ up to 20. We have checked that for width
$L=12$, lattice and finite length corrections are of order 1\%. One
example of these correlation functions can be seen in Fig.\ \ref{ss}
(with $x_1=y=0$ and $x_2=x$) on a doubly logarithmic scale. One observes
that the correlation functions nicely obey the power law
(\ref{fitsscf}), thus verifying both the correct location of the
critical point as well as the functional form of the spin-spin correlation
function in a finite strip.

We can then fit the exponent by studying the dependence with distance
of the correlation functions (\ref{fitsscf}) or by studying the dependence
with $L$ for the fixed location $x = L/2$. The first method has proven to give
smaller error bars and we obtain for the family of exponents $\eta_n$ for
$p_c = 0.1095$ and $L=20$:
\bea
\eta_{1}&=&\eta_{2}= 0.1854 \nonumber \\
\eta_{3}&=&\eta_{4}= 0.2561 \nonumber \\
\eta_{5}&=&\eta_{6}= 0.3015 \nonumber \\
\eta_{7}&=&\eta_{8}= 0.3354 \, ,
\label{valExp}
\eea
with relative errors at most of the order of 1\%.

One immediately notices two things:

{\it i)} The value for $\eta_1$ differs considerably from the value of
percolation  $\eta = 5/24 \approx 0.2083$ (see, e.g., \cite{StAh}),

{\it ii)} the exponents for higher moments are also considerably different
from $\eta_1$ which is also clear from inspection of Fig.\ \ref{ss}.

These results are compatible with the behavior of the magnetic
susceptibility that will be presented elsewhere \cite{prep}. We have also
calculated estimates for the exponents assuming different values for
$p_c$, namely, $p_c = 0.109$ and $0.110$ and the results are still distinct
from the ones of percolation (we obtain $\eta_1 = 0.180(1)$ for $p=0.109$ and
$\eta_1=0.190(1)$ for $p=0.110$). Moreover, it is only in the region very
close to $p=0.1095$ that we obtain a stable estimate for $\eta_1$ as we increase
the width $L$ of the lattices. One can then conclude from the
exponents controlling the algebraic decay of the correlation functions
that the Nishimori point is not in the percolation universality class.

Finally, we discuss the (effective) central charge $c$. It characterizes
the number of gapless degrees of freedom at the critical point and
appears as the universal coefficient of the first finite-size
correction to the free energy for periodic boundary conditions \cite{central}
\beq
f_L^{(p)} = f_{\infty}^{(p)} + {c \pi \over 6 L^2} + \ldots
\label{cDef}
\eeq
The leading term $f_{\infty}^{(p)}$ is not universal and does indeed
change already when we modify the conventions for the model (\ref{RBI}).
There are higher-order finite-size corrections to the free energy
including terms of the form $L^{-4}$.
The central charge is an important quantity identifying the CFT description
\cite{BPZ} of a fixed point. One has $c=1/2$ for the critical point of
the pure Ising model, but it has not been determined yet for the
Nishimori point.

In the process of computing $d_L$ we have also obtained estimates
of $f_L^{(p)}$ for different values of $p$. One can either fit these
values for $f_L^{(p)}$ exactly by (\ref{cDef}) ignoring further corrections
in which case the data for the smallest values of $L$ should not be used.
Or one includes a correction term of the form $L^{-4}$ which improves the
convergence with system size. These two approaches yield consistent
estimates for a given $p$. In addition, one can test that the result does
not change significantly if other higher-order corrections are added.
It should also be noted that the sensitivity of the estimates
for $c$ with respect to the location of $p_c$ is negligible in
comparison with the errors coming from the finite-size analysis.
The final result is that the following is a safe estimate for $c$
at the Nishimori point of the $\pm J$ RBIM (more details will
be given in \cite{prep}):
\beq
c = 0.464 \pm 0.004 \, .
\label{cVal}
\eeq
Assuming again the universality class of percolation,
we would expect the value for percolation in the Ising model
$c= {5 \sqrt{3} \ln{2} \over 4 \pi} \approx 0.4777$ \cite{JC}.
Even if our result (\ref{cVal}) is close to this value, it can
still be distinguished safely from percolation. This finding is
one more argument that the Nishimori point is {\em not}
in the universality class of percolation, at least the one
expected from Ising clusters.
Notice also that the central charge does not rely on a choice of observables,
and this argument can then be considered as the most general one.

The results presented in this letter provide new insight into the 2D $\pm J$
RBIM and the Nishimori point. The main result is that, according to the magnetic
exponent and the behavior of the higher moments of the correlation functions,
the universality class is different from the one of percolation, at least
considering the Ising spin variables as fundamental observables. This
conclusion is supported by the central charge which is independent of
the choice of observables and was measured here for the first time.
These results and a detailed study of the zero temperature fixed point will
provide a general and complete control of the RBIM, which one could consider
as the simplest, but most fundamental model for disordered systems in 2D.

We would like to thank S.~Franz, J.~L.~Jacobsen, J.~M.~Maillard,
G.~Mussardo, N.~Read and F.~Ritort for useful discussions and
comments.


\begin{thebibliography}{99}
\bibitem{footnote} Different variants of the phase diagram Fig.\ \ref{phaseRBI}
have been presented before, in particular in \cite{GRL} which we refer to
for further comments and references regarding global properties of
the phase diagram.
\bibitem{footnDL} We omit a factor $\beta$ since it is not important at
   finite temperature.
\end{thebibliography}
\end{document}